# Electron Photodetachment from Aqueous Anions.
# III. Dynamics of Geminate Pairs Derived from Photoexcitation of Mono- vs. Poly- atomic Anions. [1]

Rui Lian, Dmitri A. Oulianov, Robert A. Crowell, Ilya A. Shkrob [*]

*Chemistry Division , Argonne National Laboratory, Argonne, IL 60439*

Xiyi Chen, Stephen E. Bradforth

*Department of Chemistry, University of Southern California, Los Angeles, CA 90089*

**Abstract**

Photostimulated electron detachment from aqueous inorganic anions is the simplest example of solvent-mediated electron transfer. As such, this photoreaction became the subject of many ultrafast studies. Most of these studied focussed on the behavior of halide anions, in particular, iodide, that is readily accessible in the UV. In this study, we contrast the behavior of these halide anions with that of small polyatomic anions, such as pseudohalide anions (e.g., $HS^-$) and common polyvalent anions (e.g., $SO_3^{2-}$). Geminate recombination dynamics of hydrated electrons generated by 200 nm photoexcitation of aqueous anions ($I^-$, $Br^-$, $OH^-$, $HS^-$, $CNS^-$, $CO_3^{2-}$, $SO_3^{2-}$, and $Fe(CN)_6^{4-}$) have been studied. Prompt quantum yields for the formation of solvated, thermalized electrons and quantum yields for free electrons were determined. Pump-probe kinetics for 200 nm photoexcitation were compared with kinetics obtained at lower photoexcitation energy (225 nm or 242 nm) for the same anions, where possible. Free diffusion and mean force potential models of geminate recombination dynamics were used to analyze these kinetics. These analyses suggest that for polyatomic anions (including all polyvalent anions studied) the initial electron distribution has a broad component, even at relatively low photoexcitation energy. There seem to be no well-defined threshold energy below which the broadening of the distribution does not occur, as is the case for halide anions. Direct ionization to the conduction band of water is the most likely photoprocess broadening the electron distribution. The constancy of (near-unity) prompt quantum yields vs. the excitation energy as the latter is scanned across the lowest charge-transfer-to-solvent band of the anion is observed for halide anions and $Fe(CN)_6^{4-}$ but not for other anions: the prompt quantum yields are considerably less than unity and depend strongly on the excitation energy. Our study suggests that halide anions are in the class of their own; photodetachment from polyatomic, especially polyvalent, anions follows a different set of rules.

___

[*] To whom correspondence should be addressed: *Tel* 630-252-9516, *FAX* 630-2524993, *e-mail:* shkrob@anl.gov.

# 1. Introduction.

In Part I of this series,[1] quantum yields (QYs) for hydrated electrons generated by photoinduced electron detachment from various aqueous anions have been determined. Most of these anions exhibit the distinctive charge-transfer-to-solvent (CTTS) band in their UV absorption spectra that implicated the involvement of a short-lived excited state (CTTS state) mediating electron photodetachment.[2,3] Following the dissociation of this CTTS state, a geminate pair of the electron and the residual radical (such as hydroxyl for HO$^-$ and iodine atom for I$^-$) is formed in less than 200 fs. Some of these geminate pairs recombine (typically, within 0.5-1 ns), while the rest escape to the solvent bulk, yielding "free" electrons and radicals that undergo slow recombination and other secondary reactions.[4-11] The electron yields given by Sauer et al.[1] are those for *free* electrons that escaped recombination with their geminate partners. Such results *per se* are of limited import for understanding the photophysics of electron detachment in solution because this yield is a product of two factors: the prompt quantum yield for electron formation and the survival probability $\Omega_\infty$ of the resulting geminate pair.[1,4,10] The latter parameter can be determined by time-resolved measurement on a sub-nanosecond time scale. For many aqueous anions, such a measurement requires a pulsed source of ultraviolet (UV) or vacuum UV light,[2] and ultrafast pump-probe studies of these photosystems has only recently become possible.

This study pursued three objectives: (i) to extend previous observations of electron dynamics in photoexcited halide anions[4-6] and hydroxide[10] to other common CTTS anions, including polyvalent ones, (ii) to find the effect of the photon energy on these electron dynamics (extending our detailed study of iodide[6] to other anions); and (iii) to determine the prompt QY for several classes of aqueous anions. Importantly, we limit ourselves to *one-photon* excitation of these anions. Due to different selection rules, the photophysics of biphotonic excitation is different from that of monophotonic excitation (e.g., ref. 10).

To achieve these goals, recombination dynamics of hydrated electrons ($e^-_{aq}$) generated by 200 nm photoexcitation of several aqueous anions were observed and these kinetic data and photo- and actino- metric measurements were used to obtain *directly* the prompt quantum yield for the formation of thermalized electrons. Previously, such yields had been estimated for two halide anions only, I$^-$ and Br$^-$, and were found to be unity across their lower energy CTTS subband.[4,12] In this work, we present a larger pattern of these QYs. Our results suggest that polyatomic anions generally have QYs substantially lower than unity. Furthermore, some of these QYs systematically increase with the excitation energy from the very onset of the CTTS band, possibly due to the involvement of non-CTTS photoexcitation and/or side photochemical reactions.

The geminate recombination kinetics obtained for 200 nm photoexcitation of several aqueous anions were compared with pump-probe kinetics obtained for the same anions at lower photoexcitation energy (242 nm or 225 nm). Two models of geminate recombination in the CTTS photosystems, the free diffusion model [13] and the mean force potential (MFP) model,[4,10,11] were used to analyze these data sets. This analysis suggests that *for aqueous anions other than halides and simple pseudohalides (such as OH$^-$ and HS$^-$), the initial spatial distribution of photogenerated electrons has a broad component, regardless of the photoexcitation energy*. These results generalize our observations for aqueous I$^-$ [4,6] and observations of Barthel and Schwartz for Na$^-$ in ethers.[8] The narrow component in this distribution originates from the dissociation of the precursor CTTS state which proceeds through a succession of localized electron states. Such a picture has been suggested and corroborated by quantum molecular dynamics calculations carried out by several research groups.[14,15,16] It is not clear from our kinetic data whether this type of a photoprocess occurs for polyvalent anions, -- and possibly for some polyatomic monovalent anions, such as CSN$^-$.

Direct promotion of the electron from a photoexcited anion to the conduction band of the solvent is a possible cause for the occurrence of the *broad component* in the electron distribution at higher photon energy. It is well known that in photoionization of solvent and/or neutral solute molecules in polar and nonpolar liquids, the width of the electron distribution increases with the photoexcitation energy (e.g., ref. 17). This trend can be accounted for by a longer thermalization path for the conduction band electrons (emitted in the course of the direct ionization) as their kinetic energy increases. Electron photodetachment from aqueous halide anions broadly conforms to the behavior expected for this photoionization mechanism: there is a threshold energy for the electron distribution broadening (corresponding to the onset of electron promotion to the conduction band) and systematic increase in the width of the electron distribution with the total energy of photoexcitation, as observed by Bradforth and co-workers. [6]

In other cases (e.g., in low energy photoionization of water), [18,19] different mechanisms are in play. Indeed, the energetics of this photoionization are such that it can not involve these conduction band electrons; yet the resulting electron distribution is quite broad (ca. 1 nm width). Many theories of this low energy photoprocess have been suggested (see ref. 19 for a brief review). The current favorite is rapid electron transfer from photoexcited water molecule to a "pre-existing" trap in the solvent bulk (this reaction is thought to be concerted with ultrafast deprotonation of the resulting hole). [20] Our results for photo-CTTS reactions of aqueous polyatomic anions suggest that a similar photoprocess may occur in the course of electron photodetachment from such anions: there appear to exist a reaction path that yields a relatively broad initial electron distribution without the involvement of the conduction band of the solvent.

**2. Experimental.**

*200 nm photoexcitation.* The pico- and femto- second kinetic measurements reported below were obtained using a 1 kHz Ti:sapphire setup details of which are given in ref. 10. This setup provided 60 fs FWHM, 3 mJ light pulses centered at 800 nm. One part of the beam was used to generate 800 nm probe pulses while the other part was used to generate the 200 nm (fourth harmonic) pump pulses. Up to 20 µJ of the 200 nm light was produced this way (300-350 fs FWHM pulse). The pump and probe beams were perpendicularly polarized and overlapped at the surface of a 150 µm thick high-speed jet at 6.5°. Typically, 150-200 delay time points acquired on a quasi-logarithmic grid were used to obtain the decay kinetics of the electron out to 600 ps. The vertical bars in the figures represent 95% confidence limits for each data point. The details of the detection and the flow system are given elsewhere. [10,11]

*Ultrafast QY measurements (200 nm photoexcitation).* The pump power, before and after the sample, was monitored using a thermopile power meter (Ophir Optronics model 2A-SH). Normally, the concentration of the anion was such that the optical density of the solution across the jet was 2-3; the transmission of the pump pulse was therefore negligible. For obtaining the power dependence, a fast-response pyroelectric detector was used (Molectron J3-080) whose output was amplified by 20 dB using a 3-stage preamp (SRS model 240). This detector was calibrated at 1064 nm, and the readings for 200 nm light were corrected by a factor of 1.43 given by the manufacturer. These readings where 7-10% lower than the readings from the thermopile detector (both of these NIST traceable detectors had been recently calibrated by their respective manufacturers). Two methods were used to characterize radial profiles of the focussed beams at the jet surface: (i) transverse scanning of the beams by a 10 μm pinhole and (ii) imaging of the beams using a CCD camera placed in the equivalent focal plane. The typical eccentricity of these focussed beams was 1-5%. A harmonic average of $1/e$ semiaxes obtained by a 2D Gaussian fit is given below as the beam radius (eq. (A1) in the Appendix; see the Supplement). The probe beam was perfectly Gaussian (the typical profiles are shown in Fig. 1(a)), whereas the 200 nm pump beam was of inferior quality, exhibiting one or several "hot" spots. For this reason, absolute QY measurements were performed with the reverse beam geometry: the probe beam enveloped the pump beam and integrated over the imperfections in the latter. For kinetic measurements, more common geometry was used: the typical radii of the probe and pump beams where 65 μm and 210 μm, respectively, so that the pump beam enveloped the probe beam.

Absolute QY's for hydroxide and iodide were determined photometrically using an approach explained in the Appendix (see the Supplement). A power dependence of the photoinduced 800 nm absorbance from hydrated electron at delay time $t=5$ ps was plotted vs. the pump power, and QY was calculated by least squares fits using eq. (A3). For the data shown in Fig. 1(b), the beam radii of the pump and probe where 49±2 and 202±3 μm, respectively. The absolute QY obtained in this fashion depends on the assumed molar absorptivity of the electron at 800 nm (the value of 18,530 $M^{-1}$ $cm^{-1}$ [21,22] was used throughout this series), the absolute pump power (known to ±5% accuracy due to the uncertainty in the calibration of the power meters), and the radii of the pump and probe beams (known with accuracy < 3-5%). The thickness of the jet and the absorbance of photoexcitation light by the solution need not be known precisely for the QY determination.

These photometric measurements were complemented by actinometry. Electron generation in 200 nm photoexcitation of $K_4Fe(CN)_6$ was the primary actinometric standard; a prompt QY of unity was assumed for this photoreaction (section 3.2). A 1.4 μm thick film of amorphous hydrogenated silicon on a Suprasil substrate (obtained from H. Fritzsche of the University of Chicago) was used as the secondary standard. This sample was mounted on a 2D translation stage and positioned into the plane of the jet; the pump and probe beams passed through the same spot of this sample for each measurement. For each anion solution, this reference sample was inserted into the beam path immediately after the kinetic measurement. The 800 nm absorbance taken at the end of the excitation pulse was used for the normalization of $e^-_{aq}$ kinetics from these anion solutions. This 800 nm absorbance signal (ca. 0.134 OD per 1 μJ for a 59 μm $1/e$ radius pump beam, see Fig. 1S in the supplement) was linear to the highest pump fluence used for anion photoexcitation. The last step was to normalize these scaled electron kinetics so that the prompt electron yield for $Fe(CN)_6^{4-}$ was unity. For the series shown in Fig. 2, the pump power was 0.1-0.3 μJ, and Gaussian radii for the pump and probe beams were 43-46 μm and 13-20 μm, respectively.

*225 and 242 nm photoexcitation.* Kinetic experiments with 225 nm (242 nm) photoexcitation were carried out at USC using either a 200 kHz or a 1 kHz Ti:sapphire amplified laser system for which details are given elsewhere.[4,5,6] A 1 mm optical path cell with $CaF_2$ windows was used to flow the sample. Transient absorption signals down to 10 μOD can be studied using this pump-probe spectrometer. Photoexcitation pulses were generated by frequency doubling the pumped OPA output. The probe pulses were selected from the white light supercontinuum generated in a sapphire disc using a 25 nm FWHM bandpass interference filter. In dilute hydrosulfide solutions (< 0.01 M), 2-photon water ionization by 242 nm light competed with one photon electron detachment from $HS^-$ when 242 nm light irradiance was > 20 $GW/cm^2$. To avoid this complication, 0.04-0.14 M HS⁻ solutions were used and the laser irradiance was kept < 5 GW/cm$^2$. Under these conditions, two photon ionization of water was negligible, and the transient kinetics did not change with the HS⁻ concentration and pump irradiance.

**3. Results.**

*3.1. Photometric QY measurements for 200 nm photolysis of hydroxide and iodide.*

Power dependencies for the yield of hydrated electron were obtained at delay time $t=5$ ps following a 200 nm (300 fs FWHM) photoexcitation of 0.29 M KOH and 0.2 M NaI solutions (Fig. 1(b)). At this delay time, the thermalization/solvation of the electron is complete. No species other than the electron absorbed at the probe wavelength of 800 nm. The time profile of the thermalization and geminate decay kinetics did not change with the 200 nm pump power (to 2.5 μJ). The negative curvature in these power dependencies is the consequence of the beam geometry, as explained in the Appendix.

Quantum yields of 0.954±0.03 for iodide and 0.342±0.008 for hydroxide were obtained by fitting these power dependencies using eq. (A3) (the error limits indicate the standard error of these least squares fits). As explained in section 2, the actual confidence limits are probably ±10% due to the uncertainty in the assumed parameters. Near unity prompt QY for electron detachment in 220-240 nm photolysis of aqueous iodide was (indirectly) determined by Bradforth and co-workers, [4] who extrapolated the free electron yield data of Iwata et al. (for 248 nm photolysis) to $t \to 0$ using their ultrafast kinetic data.

*3.2. Actinometric QY measurements for various anions (200 nm photoexcitation).*

The concentrations of the anions studied and their molar absorptivities at 200 nm are given in Table 1 (the latter were taken from the literature sources given in Table 2 of Part I of this series). [1] Time dependencies of quantum yield $\phi(t)$ for the electron in 200 nm photoexcitation of these aqueous anions were obtained as described in section 2 and then fit between 3 ps and 600 ps by a sum of two exponentials with a nonzero offset (Fig. 2). The QYs were extrapolated to $t \to \infty$ and $t \to 0$ using these biexponential fits, and the ratio of these quantities provided an estimate for the survival probability $\Omega_\infty$ of the corresponding geminate pair (Table 1). For convenience, QYs at $t=10$ ps are also given. Note that the ratio of these actinometric QYs for hydroxide and iodide is 0.36, which is very close to the ratio of photometric QYs given in section 3.1. For comparison, free electron yields for 248 nm and 193 nm laser photoexcitation of the same anions are also given in Table 1.

Fig. 2 suggests that aqueous hexacyanoferrate(II) anion (commonly referred to as ferrocyanide) has the highest prompt QY of the aqueous anions studied in this work. The geminate decay for ferrocyanide is very slow; there is almost no decay of the electron absorbance over the first 600 ps. Bradforth and coworkers observed < 20% decrease in the hydrated electron absorbance over the first 3 ns after 220 nm photoexcitation of ferrocyanide.[23] For higher photon energy, this geminate decay may be even slower, since Sauer et al. obtained near unity QY for electron formation in 193 nm photolysis of 0.05-0.7 mM ferrocyanide.[1] A limiting quantum yield of 0.88 for 214 nm photolysis was determined by Shirom and Stein for evolution of nitrogen formed by electron scavenging in $N_2O$-saturated solutions of ferrocyanide.[24] Between 254 nm and 214 nm wavelength, this QY rapidly increased with the photon energy. The results of Sauer et al.[1] suggest that this trend continues to 193 nm. For all of these reasons, the prompt QY for ferrocyanide was taken as unity. We emphasize that in millimolar solution, ferrocyanide is associated with one or two potassium cations.[23,25] The first association constant is ca. 180 $M^{-1}$,[25] i.e., 55% and 92% of the anions are associated in 2 and 20 mM solutions, respectively. The associated anions may have slightly different photochemical properties than bare anions (see Parts I[1] and II[11] of this series). Nevertheless, no difference was observed in the recombination dynamics for 2 and 20 mM ferrocyanide solutions (see Fig. 4(b) in section 4.1).

### 3.3. Protic equilibria and secondary chemistry.

Since hydrosulfide is a strong base, hydroxide is always present in its aqueous solution. For aqueous $H_2S$ at 25°C, the first and the second ionization $pK_a$'s are 7.05 and 19, respectively,[26] and the protic equilibria involving $HS^-$ do not play a significant role in our measurements. Indeed, for 40 mM NaHS solution (Table 1) the ratio $[HS^-]/[HO^-] \approx 600$ [1,27] whereas the molar absorptivities for HS⁻ and HO⁻ at 200 nm relate as 5.6:1. Thus, the 200 nm photoexcitation of hydroxide is negligible.

Atoms (e.g., iodine) and radicals (e.g., HS) formed in the UV photolysis of monovalent anions react with these parent anions, forming dimeric species like $I_2^-$ [28] and $[HS \therefore SH]^-$. [29] The only exception to this rule is hydroxyl radical (derived from OH⁻) which deprotonates in a reaction with OH⁻ instead. [10] None of these products absorb strongly at 800 nm and their formation rates for the anion concentrations given in Table 1 are too slow to interfere with our kinetic measurements.

Three more aqueous anions, in addition to those shown in Fig. 2, were studied using 200 nm photoexcitation: chloride, perchlorate, and sulfate. Due to their weak 200 nm absorbance, 5-10 M solutions of these anions were used. These results have been discussed in Part II of this series [11] which deals with the ionic strength effect (see section 3S and Fig. 9S therein). For these photosystems, one is forced to use concentrated solutions in which this effect is very prominent. Only for sulfate does the kinetics appear to originate from photoexcitation of the nominal anion; for perchlorate and chloride the absorbance signal is likely to be from a photoexcited impurity ion. For 0.74 M sulfate, the kinetics were similar to those for sulfite (at the same ionic strength); the estimated electron yield at $t=10$ ps is ca. 0.41. [11] We also obtained the decay kinetics for electron in 0.5 M "carbonate" solution (Fig. 2S); however, due to the protic equilibrium most of the electron absorbance is actually from photoexcited hydroxide (see the caption to Fig. 2S for more details).

### 3.4. Geminate recombination: a comparison between the high and low energy excitation data.

In this section, 200 nm pump – 800 nm probe kinetics are compared with these kinetics obtained for 225 nm or 242 nm photoexcitation (Figs. 3 and 4). In the latter series the probe wavelength was 700 nm unless specified otherwise. Due to the differences in the probe wavelength and the width of the pump pulse, only kinetics at $t >$ 5 ps were compared (in the figures given below, these two sets of kinetics were normalized at 5-10 ps; e.g., Fig. 3(a)).

Fig. 3(a) demonstrates these two normalized kinetics for bromide. This anion has the first absorption maximum of CTTS band at 200 nm.[27,30,31] The 200 nm and 225 nm kinetics are exactly the same. For $(I, e_{aq}^-)$ pair, the time profile of recombination kinetics does not depend on the photon energy when it is scanned across the low energy (225 nm) CTTS subband of this anion.[4,6] Apparently, the decay kinetics for the $(Br, e_{aq}^-)$ pair follow the same pattern. In ref. 6, recombination dynamics of $(I, e_{aq}^-)$ pairs for *higher* excitation energies were studied systematically as a function of the total photon energy using *biphotonic* excitation of iodide. For this anion, electron dynamics observed after one- and two-photon excitation of the same total energy are nearly the same [6] (which is not the case for OH⁻).[10] Given these previous results, only two iodide traces are given in this study as a reference, to contrast the 200 nm and 225 nm kinetics for the iodide (Fig. 3(b)) with such kinetics for bromide (Fig. 3(a)). The 200 nm photoexcitation corresponds to the second (192 nm) CTTS subband of the iodide.[6,27,30] The 200 nm and 225 nm kinetics for I⁻ are very different: for the 200 nm photoexcitation, the survival probability is greater and the fast exponential component is less prominent. As argued in ref. 6, this transformation is consistent with broadening of the initial spatial distribution of photogenerated electrons in the course of 200 nm photoexcitation.

Fig. 3(c), traces (i) and (ii), shows 242 nm and 200 nm kinetics for hydrosulfide, respectively. For this anion, the maximum of the first CTTS band in water is at 228 nm;[32] 200 nm is close to the position of its second CTTS band ("D" band identified by Fox and Hayon [32] using a Gaussian decomposition of the UV spectra). Thus, energetically the situation is very similar to that for aqueous iodide. The change in the decay kinetics observed at the higher photoexcitation energy strikingly resembles that observed for iodide in Fig. 3(b). Except for the faster exponential component (that fully decays in 20 ps), the 242 nm kinetics for hydrosulfide look much like those observed in 200 nm photoexcitation of hydroxide (Fig. 3(c), trace (iii)) and bromide (and 242-225 nm photoexcitation of iodide), whereas the 200 nm kinetics for HS⁻ look more like those for 200 nm photoexcitation of the iodide (compare Fig. 3(b), trace (ii) and Fig. 3(c), trace (ii)).

For another pseudohalide, thiocyanate (CSN$^-$), the 225 and 200 nm kinetics look qualitatively different from those for iodide, bromide, HS$^-$, and OH$^-$ (compare Fig. 3 and 4(a)). No fast component is discernible for either photoexcitation energy; the decay of the electron is slow and the survival probability of the geminate pair is high. The latter value is ca. 20% greater for 200 nm photoexcitation than for 225 nm photoexcitation. These kinetics look more like those obtained for polyvalent anions, such as ferrocyanide (Fig. 4(b)) and sulfite (Fig. 4(c)); the only difference is the relatively low prompt quantum efficiency for the electron photodetachment from the monovalent anion.

For sulfite (Fig. 4(c)), the 200 nm and 242 nm kinetics are very similar, except for the higher escape probability for geminate pairs generated by the 200 nm photoexcitation (the survival probability at 600 ns is ca. 10% higher for the higher photon energy).

For ferrocyanide (Fig. 4(b)), the 200 nm and 225 nm kinetics ($t <$ 600 ps) are similar, save for the first 3 ps (i.e., the thermalization phase); the latter difference originates through the difference in the probe wavelength used to observe the electron dynamics (800 nm and 500 nm, respectively).

For all of these anions, the survival probability either increases with the photoexcitation energy or, at least, stays the same. A systematic increase in $\Omega_\infty$ with increasing photoexcitation energy has already been observed by Schwartz and coworkers for sodide in THF.[8] Bradforth and coworkers observed an increase in the survival probability with the *total* excitation energy of biphotonic photoexcitation of aqueous iodide.[6] Apparently, this trend pertains to all CTTS anions regardless of the solvent polarity and/or photoexcitation mode. The increase in the survival probability with increasing photoexcitation energy partially (but not entirely) accounts for the increase in the free electron yield that is seen in the data of Table 1.

**4. Discussion.**

*4.1. Simulation of kinetic traces for polyvalent anions.*

For polyvalent anions and CNS⁻, a free diffusion model complemented with a prescribed initial distribution $P(r_i)$ of the electrons around their parent radicals provides a good recipe for simulating geminate pair dynamics. Hereafter, we assume spherically symmetrical distribution $P(r_i)$ of the electrons and the mean force potential $U(r) = kT\, u(r)$ for the geminate partners. In the free diffusion model,[13] the interaction between these geminate partners is neglected (i.e., $u(r) = 0$), and the pair decays via diffusion-controlled recombination at the contact radius $r = d$. The survival probability $\Omega(t)$ for a pair generated at a distance $r = r_i$ is given by

$$\Omega(t) = 1 - (d/r_i)\, \text{erfc}\left[(r_i - d)/2\sqrt{Dt}\right] \qquad (1)$$

where $D$ is the mutual diffusion coefficient. The survival probability of the geminate pairs is given by $\Omega_\infty = 1 - \langle d/r_i \rangle$, which is a function of the initial distribution $P(r_i)$ of the electron only. Equation (1) is averaged over this initial distribution. The empirical distribution $P(r_i)$ that gave the best results was an $r^2$-exponential distribution with average width $\langle r_i \rangle = 3 b_E$,

$$4\pi r_i^2 P(r_i) = \left(2 b_E^3\right)^{-1} r_i^2\, \exp(-r_i/b_E) \qquad (2)$$

An $r^2$-Gaussian distribution, for which $\langle r_i \rangle = 2\pi^{-1/2} b_G$,

$$4\pi r_i^2 P(r_i) = 4\pi^{-1/2}\, b_G^{-3}\, r_i^2\, \exp\left(-[r_i/b_G]^2\right) \qquad (3)$$

also provided good fits to these data. In these simulations, we assumed that $d=0.5$ nm and $D=4.5\times10^{-5}$ cm$^2$/s (which is the diffusion coefficient for hydrated electron in pure water). Only $t>5$ ps kinetics were used to obtain the least squares fit. A typical fit for 200 nm photoexcitation of sulfite is shown in Fig. 4(c). The following optimum parameters were obtained: $b_G=1.07\pm0.03$ nm ($\langle r_i \rangle=1.27$ nm) and $\Omega_\infty=0.545$ (for the $r^2$-Gaussian distribution) and $b_E=0.39\pm0.01$ nm ($\langle r_i \rangle=1.3$ nm) and $\Omega_\infty=0.53$ (for the $r^2$-exponential distribution). For 242 nm photoexcitation, $b_E=0.32\pm0.04$ nm ($\langle r_i \rangle=1.1$ nm) and $\Omega_\infty=0.47$ (for the $r^2$-exponential distribution); i.e., the escape probability is ca. 13% higher for the 200 nm photoexcitation. Note that this increase is clearly insufficient to account for a two times increase in the free electron yield between 248 nm and 200 nm (Table 1) for sulfite. Apparently, for this anion the *prompt* quantum yield of the electron *increases* with the photon energy.

For thiocyanate anion excited by 200 nm and 225 nm photons, we obtained $b_E$=0.61±0.08 nm ($\langle r_i \rangle$=1.92 nm) and $\Omega_\infty$=0.655 and $b_E$=0.26±0.09 nm ($\langle r_i \rangle$=0.98 nm) and $\Omega_\infty$=0.414, respectively (for the $r^2$-exponential distribution). Thus, the escape probability increases by 60% from 242 nm to 200 nm photoexcitation, whereas the free electron yield increases more than 10 times from 248 nm to 200 nm (Table 1). Once more, the dramatic increase in the free electron QY can only be due the increase in the *prompt* electron yield rather than the survival probability of the geminate pair.

For ferrocyanide, geminate recombination is so slow and inefficient that only a lower estimate for $\langle r_i \rangle$ can be obtained, ca. 0.31 nm (which would correspond to $\Omega_\infty$=0.66). For all of these anions, the average ejection distances $\langle r_i \rangle$ are over 1 nm, which is the typical electron - hole separation in the low-energy excitation of neat water (section 1). This suggests that the electron promoted from photoexcited polyvalent anions is ejected directly into the conduction band of the solvent.

Hydrosulfide is a monovalent anion for which the geminate dynamics observed following 200 nm photoexcitation may, on principle, be simulated using eq. (1) and an *ad hoc* electron distribution $P(r_i)$ shown in Fig. 2S in the supplement (Note that for monovalent anions other than thiocyanate, no distribution $P(r_i)$ can be found to account for the kinetics observed, provided that the diffusion is free). This distribution was determined by the least squares optimization of probability weights. The optimum distribution consists of a narrow peak at $r_i < 1$ nm (that includes 84% of the electrons with $\langle r_i \rangle$ of 0.63 nm) and a broad peak centered at 2.4 nm (which includes 16% of the electrons with $\langle r_i \rangle$ of 3 nm). This simulation suggests that the electron distribution for HS⁻ photoexcited by 200 nm light might be at least as broad as that for CSN⁻. Note that for HS⁻ excited by 242 nm and 200 nm photons, the escape probability of the electron (as estimated from biexponential fits) changes with the photon energy ca. 3 times, whereas the free electron yield changes by an order of magnitude from 248 nm to 200 nm (Table 1). This anion provides yet another example of an anion photosystem in which the prompt QY of photoelectron dramatically increases with the photoexcitation energy.

*4.2. Simulation of kinetic traces for monovalent anions.*

The approach used to simulate kinetics for geminate pairs derived from monovalent anions is based on Shushin's semianalytical theory [6,10,11,33] for diffusion controlled reactions in a potential well. In this model, geminate partners interact by means of an attractive mean force potential (MFP), $U(r) = kT\, u(r)$, with Onsager radius $a$ (at which $u(a) = -1$). It can be shown that the recombination and escape from the potential well are pseudo-first-order reactions; the corresponding constants $W_r$ and $W_d$ can be evaluated in a rather complex way from the radial profile of the MFP and the diffusion constant $D$.[10,33] The survival probability $\Omega(t)$ is given by [10]

$$\Omega(t) \approx 1 - (1-p_d)\left[1 + \frac{\mathrm{Im}\ \lambda^{-1}\exp(\lambda^2 Wt)\ \mathrm{erfc}(\lambda\sqrt{Wt})}{\mathrm{Im}\ \lambda}\right] \qquad (4)$$

where $W = W_r + W_d$ is the total decay rate, $p_d = W_d/W$ is the escape probability of a geminate pair generated in the potential well ($r_i < a$), and $\lambda = \alpha/2 + i(1-\alpha^2/4)^{1/2}$, where $\alpha = p_d\sqrt{a^2 W/D}$ is a dimensionless parameter. On a short time scale ($Wt \ll 1$), the survival probability $\Omega(t)$ decays exponentially as $\exp(-Wt)$, whereas on a long time scale ($Wt \gg 1$) it decays by a power law as $p_d(1 + a/\sqrt{\pi t D})$, asymptotically approaching $p_d$ at $t \to \infty$. In the derivation of eq. (4), it was assumed that all of the geminate pairs were generated inside the potential well (that is, $r_i < a$). In a more general case, Sushin's equations must be averaged over the initial electron distribution $P(r_i)$ of the electrons; a compact analytical expression similar to eq. (4) cannot be obtained in such a case but a numerical solution can be computed from the Laplace transform of $\Omega(t)$ using eqs. (B11), and (B24) given in Appendix B of ref. 10. The escape probability $\Omega_\infty$ for an arbitrary electron distribution is given by

$$\Omega_\infty = 1 - (1-p_d)\langle \min(1, a/r_i)\rangle \tag{5}$$

(see eq. (B25) in ref. 10). For an $r^2$-exponential distribution, eq. (5) simplifies to

$$\Omega_\infty = p_d + (1-p_d)\exp(-\upsilon)[1 + \upsilon/2] \tag{6}$$

(see eq. (B28) in ref. 10), where another dimensionless parameter, $\upsilon = 3a/\langle r_i\rangle$ is introduced. For $\upsilon \gg 1$, $\Omega_\infty = p_d$ and $\Omega(t)$ is given by eq. (4). Note that in Shushin's theory all geminate pairs for which $r_i < a$ have the same recombination dynamics, so that the overall kinetics are not sensitive to the exact profile of $P(r_i)$ at these short distances.

For halide anions, I⁻ and Br⁻, the kinetics do not change with the photoexcitation energy across the low energy CTTS subband, and we assume that the electron distribution does not change with this energy. (It may also be assumed that in this energy regime $\upsilon \gg 1$, and the geminate kinetics are not sensitive to the changes in the electron distribution for $r_i < a$). The optimum fit parameters are given in Table 2. The model parameters are similar for both of these halide anions, with a greater escape probability $p_d$, residence time $W^{-1}$ and estimated Onsager radius $a$ for bromide (0.58 vs. 0.42 nm; these Onsager radii were estimated assuming a value of $4.5 \times 10^{-5}$ cm$^2$/s for the diffusion coefficient $D$). These parameters are close to those obtained for hydroxide (Table 2); the the Onsager radius $a$ for the $(OH, e_{aq}^-)$ pairs is ca. 0.53 nm. As shown in ref. 6, for aqueous iodide excited by photons with total energy > 6 eV, the kinetics can still be fit using the same MFP parameters assuming that the electron distribution exhibits a broad, diffuse component in addition to the narrow ($\upsilon \gg 1$) distribution typical of the low-energy CTTS photoexcitation. The broad and the narrow components correspond to the direct ionization and CTTS state mediated electron detachment, respectively. A monomodal (e.g., $r^2$-exponential) distribution with $b_E$ increasing smoothly with the photon energy provided a rather poor overall fit for the kinetics;[6] the best quality multi-trace fit was obtained by letting the weights of the narrow ($r_i < a$) and broad ($r_i > a$) distribution be floating parameters as a function of the excitation energy. To fit the 200 nm data for the iodide, such a bimodal distribution is not needed: it is possible to fit the kinetic data using the $r^2$-exponential distribution with $\langle r_i\rangle \approx 0.57$ nm which corresponds to $\upsilon \approx 2.2$. The choice between the bimodal and monomodal distribution is difficult to make since the shape of this prescribed monomodal distribution is arbitrary. The advantage of using the bimodal distribution is that the two postulated components have clear physical meaning. The disadvantage is that extra assumptions have to be made to reduce the parameter space (e.g., in ref. 6 we assumed that the broad distribution has the same width at different energies, which is not self-evident, etc.) Such a dilemma is always present in the analysis of recombination kinetics due to the relatively weak dependence of $\Omega(t)$ on the fine details of the $P(r_i)$ profile.

The broad initial electron distribution is also needed to account for the HS$^-$ kinetics given in Fig. 5(b). Following the approach used for iodide [4,6] and hydroxide, [10] we assumed that the lowest-energy (242 nm) kinetics correspond to the case when $\upsilon >> 1$ and then optimized parameter $\upsilon$; this procedure gave $\upsilon \approx 3$ for 225 nm photoexcitation and $\upsilon \approx 2.1$ for 200 nm photoexcitation. Note that for hydrosulfide the life time of caged pairs (parameter $W^{-1}$) is 3-5 times shorter than that for OH$^-$ and halide anions (Table 2); the Onsager radius is also smaller (ca. 0.32 nm). The probability of escape, $p_d$, at the onset of the CTTS band, increases in the order $Br^- > I^- > OH^- > HS^-$.

### 4.3. Synopsis.

The following general pattern emerges from our results: Most photoexcited aqueous anions yield a broad initial distribution of electron-radical separations, at least under some excitation conditions. Photoexcitation of polyvalent anions as well as some pseudohalide anions, such as CNS$^-$, always yields this type of electron distribution, regardless of the photon energy. For halide anions, the electron distribution changes from a narrow one to a broad one when the photon energy exceeds a certain threshold value. [4,6] The QY data suggest a systematic broadening of the electron distribution with the excitation energy (e.g., for SO$_3^{2-}$, CSN$^-$, HS$^-$, and I$^-$) that accounts for the increase in the survival probability and the free electron yield. For iodide in water and sodide in THF, this increase has already been reported. [6,8] The prompt QY for aqueous halides is near unity regardless of the photoexcitation energy, but this is not the case for other aqueous anions. For thiocyanate, sulfite, and hydrosulfide the increase in the free electron QY with increasing photon energy is much greater than the concomitant increase in the escape probability of the electron: for these anions, the *prompt* QY definitely *increases* with the photoexcitation energy. For 200 nm photoexcitation, prompt QYs for pseudohalides are 30-40% of the QYs for halides; the prompt QYs for polyvalent anions do not follow any obvious pattern. Some trends in these QYs and possible explanations for these trends were discussed in Part I of this series. [1]

The current thinking is that the broad electron distribution originates mainly through the direct photoionization when the electron is promoted into the conduction band of water. For some anions (e.g., ferrocyanide), [34] the ejection distances are so long that no other mechanisms seem possible. The broadening of the electron distribution with increasing excitation energy may be viewed as a competition between this direct ionization (which involves extended state electrons) and dissociation of the precursor CTTS state that involves localized electrons only, as explained in the Introduction. The latter photoprocess prevails at low excitation energies since the conduction band is inaccessible at these low energies.

Our results do not wholly agree with this picture: for some of these anions (both mono- and poly- valent) the broadening of the electron distribution with increasing excitation energy occurs across their entire lower CTTS band. For polyvalent anions (and even some monovalent anions, such as thiocyanate), it is not clear from our kinetic data that the CTTS state dissociation analogous to that observed for halides occurs at all. Energetics alone do not account for the remarkable differences observed between the picosecond dynamics of the corresponding geminate pairs. For example, $HS^-$, $I^-$, $CSN^-$, and $SO_3^{2-}$ all have similar CTTS band energies and photoelectron emission threshold energies (see Table 1 in ref. 35), yet the latter two anions yield broad electron distributions regardless of the photoexcitation energy whereas iodide and hydrosulfide yield a narrow electron distribution when excited by low energy photons and a broad electron distribution when excited by high energy photons.

**5. Conclusion.**

Time-dependent photoelectron quantum yields generated by 200 nm photoexcitation of several aqueous anions have been obtained. These kinetic profiles were compared with the 225 nm (and/or 242 nm) kinetic data for the same anions. It is shown that electron photodetachment from molecular anions follows a different set of rules than electron photodetachment from halide anions. In particular, the distribution of electrons promoted from non-halide anions (including HS$^-$, CNS$^-$ and all polyvalent anions) shows a broad component from the very onset of the lower CTTS band. For halide anions there is a well-defined threshold energy below which only narrow electron distributions are obtained. It is not presently clear what specific mechanisms are responsible for the observed electron distributions.

Previously, we speculated that near unity prompt quantum yield for electron promoted from the aqueous CTTS anions might be the general feature of electron photodetachment from such anions.[4] This study shows that pseudohalide anions such as HO$^-$ and HS$^-$ (for which there are no known side photoreactions or discernible bound-to-bound absorbances) exhibit substantially lower prompt QYs for electron detachment than the halides (Table 1). Another peculiarity of electron photodetachment from small *polyatomic* anions is that the prompt QY across the lower CTTS band is not constant, as is apparently the case for halide anions.

From the outset of photochemical studies of CTTS anions, it has been believed that studies of electron photodetachment from aqueous halides, such as iodide, would provide critical insight into the analogous photoreactions for more complex polyatomic anions. This work suggests that even the simplest polyatomic anions behave somewhat differently from these halides. More experimental and theoretical studies are therefore needed before any generalization is possible. In particular, theoretical modeling of CTTS state dissociation for polyatomic, especially polyvalent, anions would be beneficial .since presently such models exist for halide anions only. As the latter anions appear to be in a class of their own, these models do not shed much light onto the general case.

**6. Acknowledgement.**

We thank Drs. D. M. Bartels and S. Pimblott of NDRL, Prof. B. J. Schwartz of UCLA, Drs. M. C. Sauer, Jr. and C. D. Jonah of ANL, and Dr. S. V. Lymar of BNL for many useful discussions. The research at the ANL was supported by the Office of Science, Division of Chemical Sciences, US-DOE under contract number W-31-109-ENG-38. The research at USC was supported by the National Science Foundation and the David and Lucile Packard Foundation. S. E. B. is a Cotrell Scholar of the Research Corporation.

***Supporting Information Available:*** A PDF file containing the Appendix and Figs. 1S to 3S with captions. This material is available free of charge via the Internet at http://pubs.acs.org.

**Table 1.**

**Actinometric estimates for quantum yields of hydrated electron in 200 nm photolysis of aqueous anions at 23°C.**

| anion | a | $\varepsilon_{200}$ [b] | $t \to 0$ | $t = 10$ ps | $t \to \infty$ | $\Omega_\infty$,[c] x 100 | QY[d] 248 nm | QY[d] 193 nm |
|---|---|---|---|---|---|---|---|---|
| I$^-$ | 20 | 10 | 0.901 | 0.715 | 0.415 | 46.1 | 0.286 | 0.497 |
| Br$^-$ | 20 | 10 | 0.906 | 0.697 | 0.292 | 31.8 | - | 0.365 |
| HO$^-$ | 200 | 1 | 0.378 | 0.261 | 0.092 | 24.5 | - | 0.112 |
| HS$^-$ | 40 | 5.6 | 0.329 | 0.214 | 0.139 | 42.2 | 0.014 | 0.298 |
| CNS$^-$ | 25 | 0.8-1 | 0.304 | 0.282 | 0.231 | 76.2 | 0.019 | 0.306 |
| SO$_3^{2-}$ | 40 | 6.3 | 0.353 | 0.313 | 0.231 | 65.4 | 0.108 | 0.391 |
| [KFe(CN)$_6$]$^{3-}$ | 20 | 13.07 | 1.0 | 0.995 | 0.963 | >96 | 0.674 | 1.018 |

a) millimolar concentration of the anion in the photolysate;

b) molar absorptivity of the anion at 200 nm, in $10^3$ M$^{-1}$ cm$^{-1}$;

c) survival probability of the geminate electron/radical pair;

d) quantum yields for free electron formation in 248 nm and 193 nm photolysis (from ref. 1).

**Table 2.**

**Simulation parameters for geminate pairs derived from monovalent anions.**

| anion | $\lambda^*$, nm [a] | $p_d$ | $\alpha$ [b] | $W^{-1}$, ps | $\upsilon$ |
|---|---|---|---|---|---|
| I⁻ | 225 | 0.216 | 0.34 | 14.0 | c |
|    | 200 |       |      |      | 2.17 |
| Br⁻ | 200 & 225 | 0.268 | 0.57 | 16.5 | c |
| HO⁻ | 200 | 0.173 | 0.375 | 12.0 | c |
| HS⁻ | 242 |       |       |     | c |
|     | 225 | 0.121 | 0.306 | 3.5 | 3.03 |
|     | 200 |       |       |     | 2.14 |

a) photoexcitation wavelength;

b) dimensionless parameter $\alpha = p_d \sqrt{a^2 W/D}$;

c) $\upsilon \gg 1$.

**References.**

**Figure captions.**

**Fig. 1.**

(a) 200 nm pump *(filled circles)* and 800 nm probe *(filled squares)* beam profiles for the photometric experiment in (b). The *1/e* radii of these two laser beams were 49±2 and 202±3 μm, respectively. (b) Photoinduced absorbance (800 nm) observed at *t*=5 ps in 200 nm laser photolysis of aqueous solutions of (i) 0.2 M NaI *(open circles)* and (ii) 0.29 M KOH *(open squares)*. Solid lines are least square fits to eq. (A3). This 800 nm absorbance signal is from thermalized photoelectrons detached from their parent anions. See sections 2 and 3.1 for more detail.

**Fig. 2.**

Time-dependent quantum yield $\phi(t)$ of photoelectrons vs. delay time *t* of 800 nm probe pulse. See section 3.2 for more detail. Single 200 nm photon excitation of aqueous (i) iodide *(open circles),* (ii) bromide *(filled squares),* (iii) hydroxide *(open squares),* (iv) hydrosulfide *(open triangles),* (v) thiocyanate *(open diamonds),* (vii) sulfite *(filled triangles),* and (viii) ferrocyanide *(filled diamonds)*. The anion concentrations are given in the first column of Table 1.

**Fig. 3.**

Normalized transient absorbance signals from thermalized photoelectrons generated in laser photolysis of aqueous (a) bromide, (b) iodide, and (c) hydrosulfide *(i,ii)* and hydroxide *(iii)*. The symbols indicate the kinetics obtained by 200 nm photoexcitation of these anions, noisy *(yellow)* lines indicate the kinetics obtained by 225 nm (a,b) or 242 nm (c) photoexcitation. *(Black)* solid lines indicate the least squares fits of these kinetics using eq. (4) in section 4.2. In (a,b), the 200 nm kinetics obtained using 800 nm *(squares)* and 1000 nm *(circles)* probe light are shown in the same plot, to illustrate the invariance of the decay kinetics of thermalized electrons vs. the probe wavelength.

**Fig. 4.**

Same as Fig. 3, for *(a)* thiocyanate, *(b)* ferrocyanide, and *(c)* sulfite. *(Black)* solid lines are least squares fits obtained using the free diffusion model given in section 4.1 assuming the $r^2$-exponential distribution of photoelectrons. In *(b)*, kinetics obtained by 200 nm photoexcitation of 2 mM *(open squares)* and 20 mM *(open circles)* ferrocyanide are given in the same plot, for comparison.

**Fig. 5.**

Time-dependent survival probability $\Omega(t)$ of the geminate pairs vs. reduced time $Wt$ (section 4.2). The simulation parameters obtained from least squares fits to the generalized Shushin's model equations are given in Table 2. The geminate $\left(X, e_{aq}^{-}\right)$ pairs are generated by *(a)* 225 nm *(i)* and 200 nm *(ii)* photoexcitation of aqueous iodide and *(b)* 242 nm *(i)*, 225 nm *(ii),* and 200 nm *(iii)* photoexcitation of aqueous hydrosulfide.

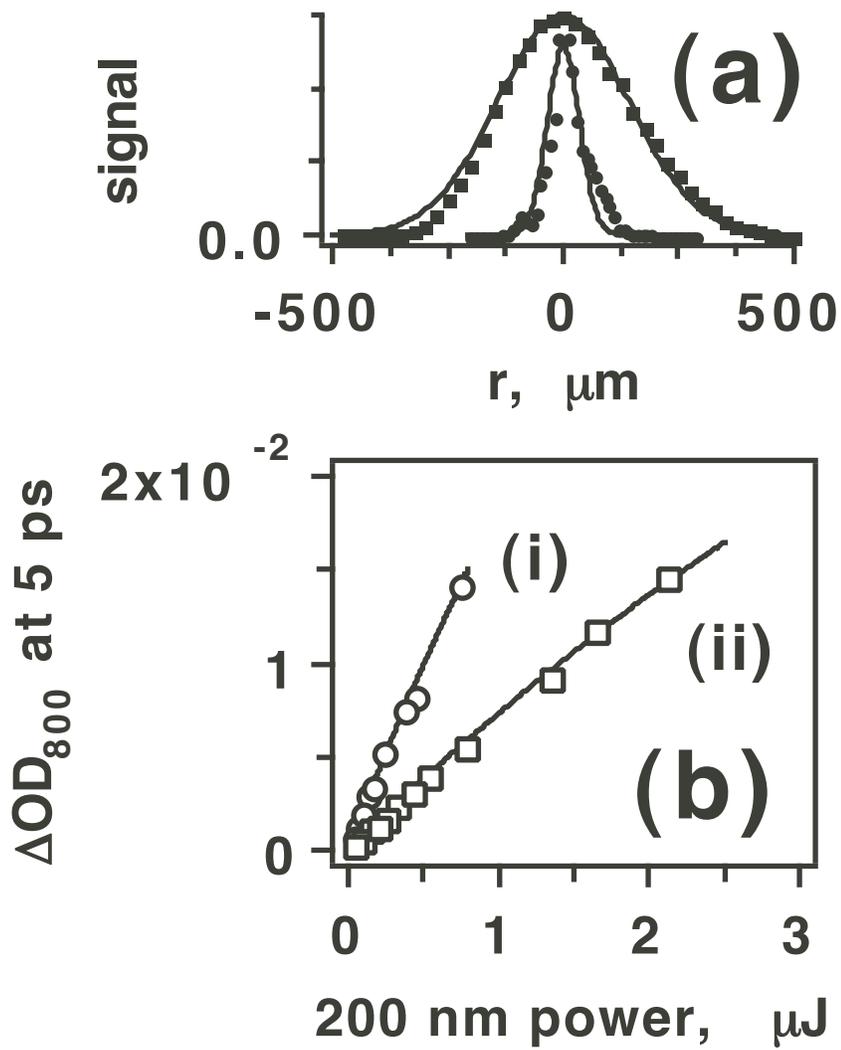

Lian et al.; Fig. 1

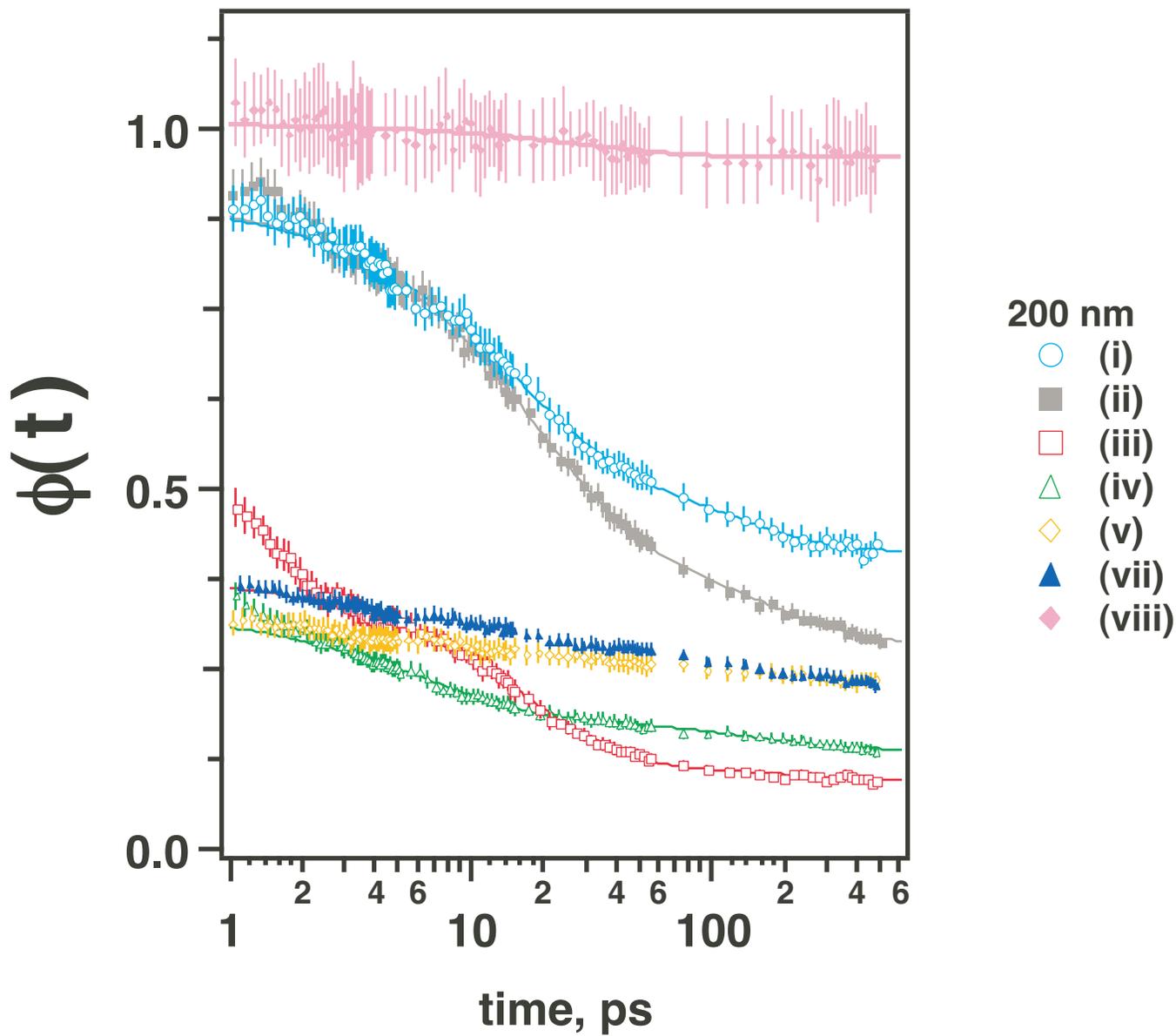

Lian et al.; Fig. 2

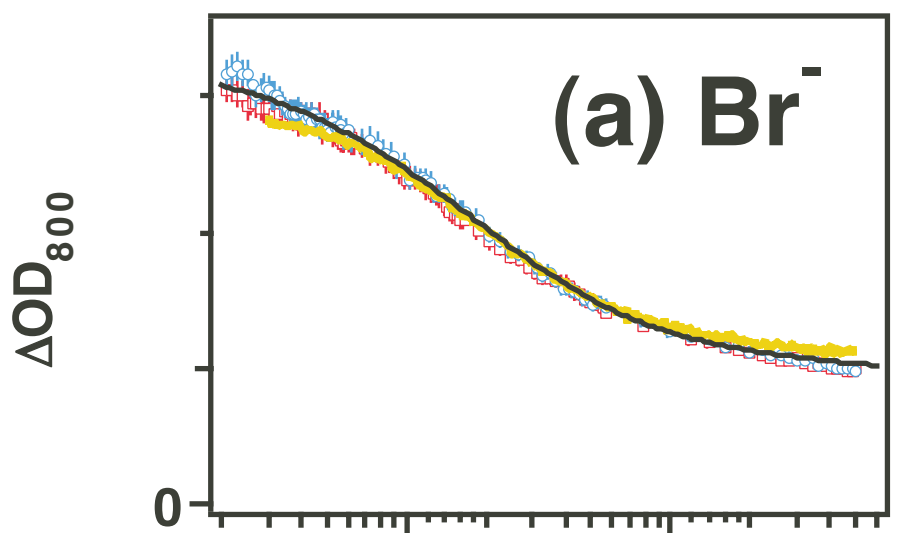
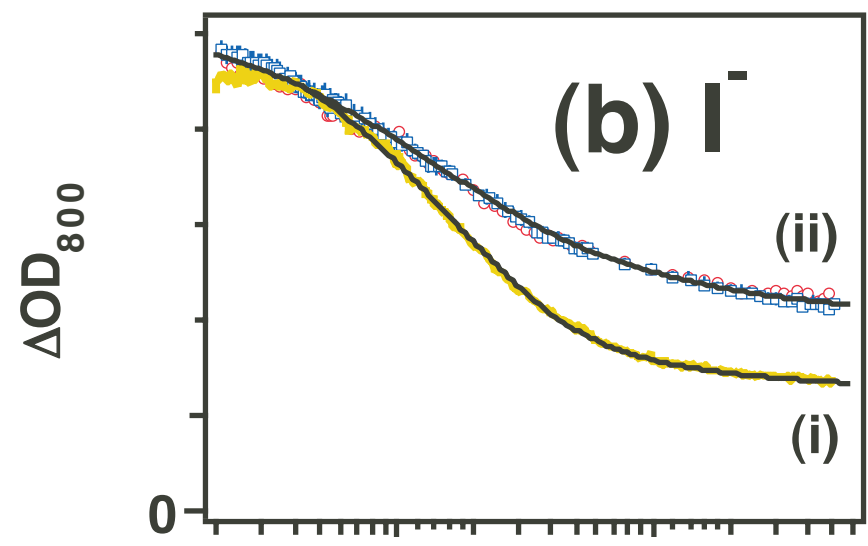
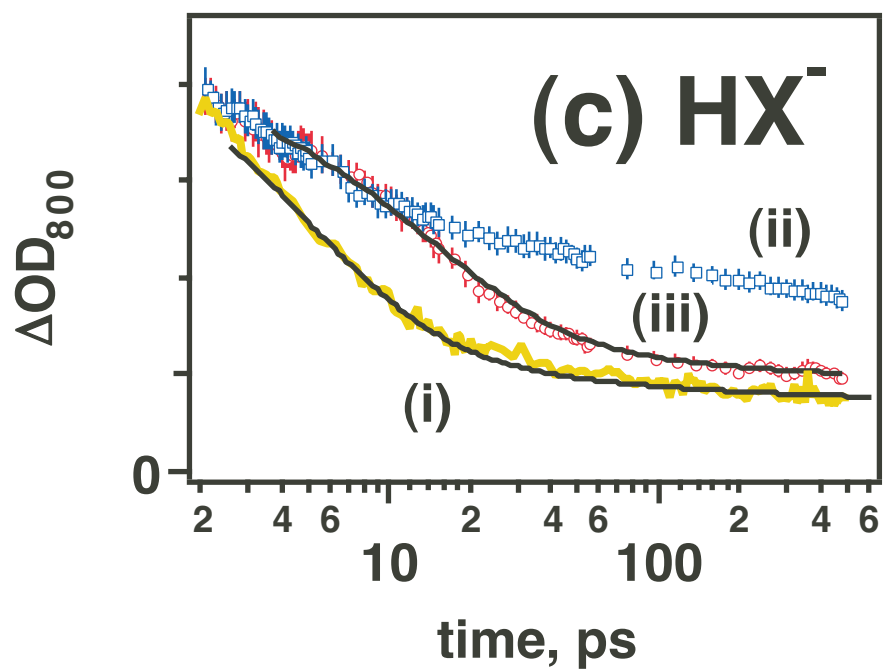

Lian et al.; Fig. 3

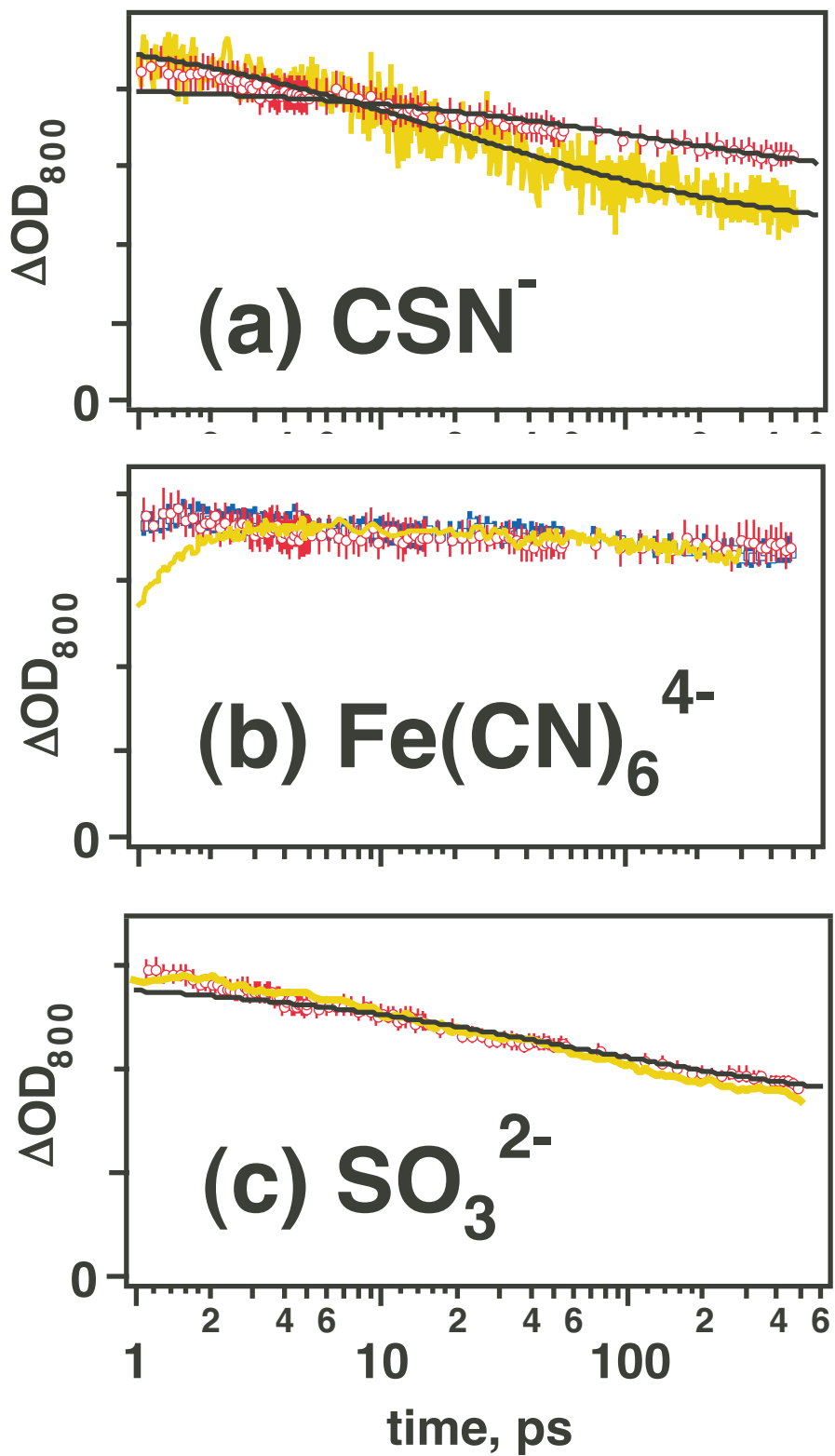

Lian et al.; Fig. 4

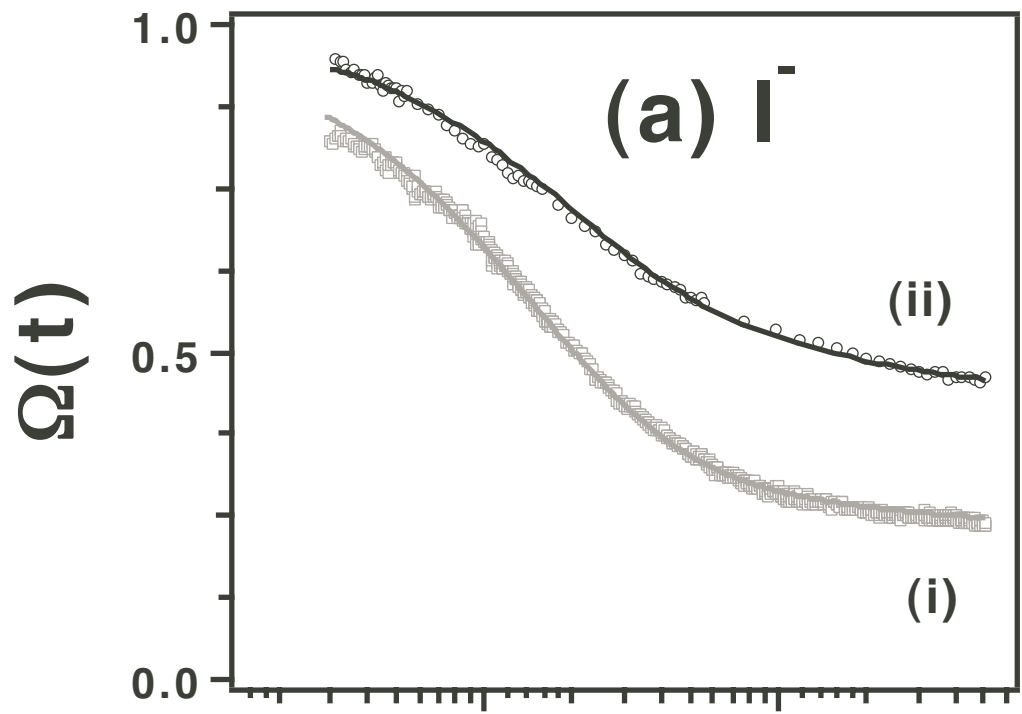
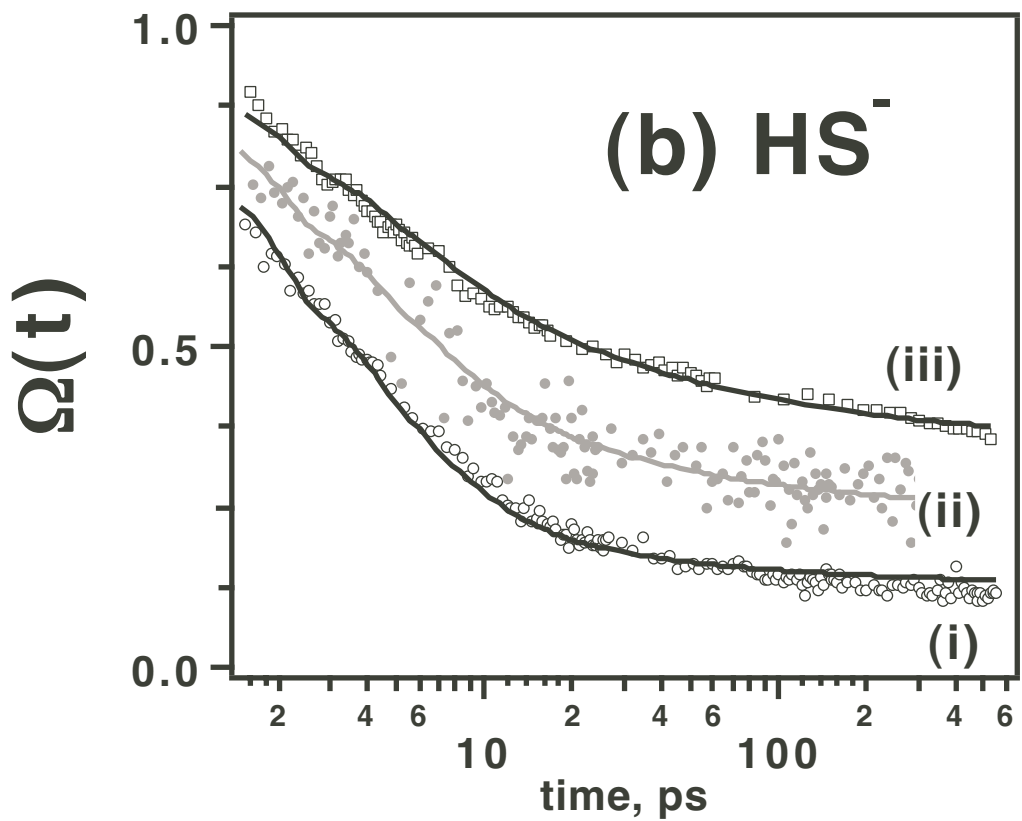

Lian et al.; Fig. 5

**SUPPLEMENTARY MATERIAL**  jp0000000

*Journal of Physical Chemistry A, Received * 2004*

**Supporting Information.**

**Appendix: Equations for the absolute QY measurement.**

To estimate the QY, the following approach was used: The (axially symmetrical) radial profiles for the pump and probe beams were assumed to be Gaussian, so that the incident (time-integrated) photon fluence $J_{pump}(r)$ for the pump beam was given by

$$J_{pump}(r) = J_0 \; \exp(-[r/r_{pump}]^2) \tag{A1}$$

and the beam power was $I_{pump} = \pi r_{pump}^2 J_0$. As the pump beam penetrates the sample, it is attenuated as $\exp(-\beta x)$, where $\beta$ is the absorption coefficient and $x$ is the penetration depth of the excitation light. Neglecting the absorbance of 200 nm light by photogenerated species and assuming that the optical density of a sample of thickness $L$ is much greater than unity ($\beta L \gg 1$), the total number of photons absorbed by the sample of thickness $L$ is given by $I_{pump}$. The average concentration $P(r)$ of the photoproduct(s) across the sample is given by $P(r) = \phi J_{pump}(r)/L$, where $\phi$ is the quantum yield for the photoproduct(s). Let us introduce $\beta_{probe}$, the molar absorptivity of the photoproduct(s) at the probe wavelength. The relative loss in the transmission of the probe light is given by

$$\left(-\frac{\Delta T}{T}\right)_{probe} = \int_0^\infty dr \; r \; J_{probe}(r) \left[1 - \exp(-\beta_{probe} L P(r))\right] \Big/ \int_0^\infty dr \; r \; J_{probe}(r) \tag{A2}$$

where $J_{probe}(r)$ is the radial distribution of the probe beam given by eq. (A1) with the beam radius of $r_{probe}$. Combining all of these equations and introducing dimensionless parameters $\xi = (r/r_{probe})^2$, $\chi = (r_{pump}/r_{probe})^2$, and $\delta = \beta_{probe} \phi \, I_{pump}/r_{pump}^2$ we obtain

$$\left(-\frac{\Delta T}{T}\right)_{probe} = \int_0^\infty d\xi \; e^{-\xi} \left[1 - \exp(-\delta e^{-\xi/\chi})\right] \tag{A3}$$

For small $\delta$, the exponential function in the square brackets may be expanded into a series and the integral taken analytically,

$$(-\Delta T/T)_{probe} = \chi \sum_{n=1}^\infty (-1)^{n-1} \delta^n / n! (n+\chi) \approx \chi \left[\frac{\delta}{1+\chi} - \frac{\delta^2}{2(2+\chi)}\right] \tag{A4}$$

Thus, the dependence of $(-\Delta T/T)_{probe}$ as a function of laser power (which is factoried into parameter $\delta$) has a negative curvature. Solving eq. (A4) for $\chi \ll 1$ (which corresponds to the probe beam enveloping the pump beam), one obtains a general formula

$$\phi \approx \frac{\pi(r_{pump}^2 + r_{probe}^2)}{\beta_{probe} I_{pump}} \left(\frac{-\Delta T}{T}\right)_{probe} \left[1 + \frac{1}{4}\left(\frac{r_{probe}}{r_{pump}}\right)^2 \left(\frac{-\Delta T}{T}\right)_{probe}\right] \quad (A5)$$

which is correct when the second term in the square brackets is much less than unity. For $\chi \gg 1$ (the pump beam enveloping the probe beam), this term should be omitted. If the correction term is not small, integration of eq. (A3) must be carried out numerically.

**Captions to figures 1S to 3S.**

**Fig. 1S.**

(a) The typical transient absorbance signal from a 1.4 μm thick film of undoped *a*-Si:H (8 at. % H) on 1 mm silica substrate *(open circles)*. This absorbance was detected at 800 nm, where both free conduction band electrons (short-lived signal) and trapped charges in the band tails (long-lived signal) absorb the probe light. This sample was obtained from Prof. H. Fritzsche and Dr. P. Stradins of the James Frank Institute, University of Chicago. The details of the sample preparation are given by Shkrob, I. A.; Crowell, R. A. *Phys. Rev. B* **1998**, *57*, 12207. The solid line is a convolution of a Gaussian profile of the 200 nm pump pulse (380 fs FWHM) and the biexponential decay kinetics. The maximum *ΔOD* signal is plotted vs. the incident power of 200 nm beam (1/*e* radius of 59 μm) in (b) *(empty circles)*. In this plot, 300 nJ corresponds to the photon fluence of 2.7 mJ/cm$^2$. The slope of the linear dependence *(solid line)* is 0.134 per mJ. This thin film sample was used as a convenient actinometric standard, as explained in section 2. The 200 nm kinetics shown in Figs. 2, 3, and 4 were obtained using 200 nm laser pulses of 100-300 nJ.

**Fig. 2S.**

Time-dependent quantum yield $\phi(t)$ of photoelectrons for 200 nm photoexcitation of 0.5 M carbonate solution (open circles). The vertical bars are 95% confidence limits; the line drawn through the points is a biexponential fit. The extrapolated QYs at zero time and infinity are 0.724 and 0.262, respectively (19% escape fraction). The free-electron QY for $CO_3^{2-}$ at 193 nm is 0.453.[1] For carbonic acid, the $pK_a$'s are 6.35 and 10.33,[26] and the presence of hydroxide in the carbonate solutions cannot be neglected. For molar concentration $c$ of the carbonate > 1 mM, $[OH^-]/[CO_3^{2-}] \approx 0.015/\sqrt{c}$, so that 0.5 M carbonate solution contains 10 mM of $OH^-$ and $HCO_3^-$ (see Fig. 8S in ref. 1). Though the molar absorptivity of $CO_3^{2-}$ at 200 nm is not known, it is lower than its 193 nm value, which is ca. 150 M$^{-1}$ cm$^{-1}$.[1] At 193 nm, the molar absorptivities of bicarbonate and carbonate anions are similar; if this similarity also holds for longer wavelengths, the molar absorptivity of the carbonate is ca. 30 M$^{-1}$ cm$^{-1}$.[27] This value is much lower than 1000 M$^{-1}$ cm$^{-1}$ for hydroxide (Table 1). Using these estimates, simple calculation suggests that the 200 nm light is absorbed in a 3:2 ratio by the carbonate and hydroxide anions, respectively. Therefore the estimate for the prompt QY for the "carbonate" given above should be increased from 0.72 to ca. 0.95. Note that despite this large prompt QY, free electron yields for $CO_3^{2-}$ and $OH^-$ are comparable, i.e., the survival probability of the $(CO_3^-, e_{aq}^-)$ pair is relatively low. The probable cause for this inefficient escape is the ionic strength effect which is considered in more detail in ref. 11. Lowering the concentration of carbonate (to reduce this complicating effect) was impractical since in dilute carbonate solutions an even lower fraction of 200 nm light would be absorbed by $CO_3^{2-}$. Buffering the *pH* of the solution was also impractical since the buffer salts (such as phosphates and borates) would absorb the photoexcitation light.

**Fig. 3S.**

Fitting the 200 nm excitation kinetics for aqueous hydrosulfide *(open circles)* in the (prescribed) free diffusion model with an arbitrary initial spatial distribution $4\pi r_i^2 P(r_i)$ of photoelectrons. The (discrete) weights of this distribution (a) were obtained by a least squares optimization; the fit curve is shown in (b). Note the logarithmic scales.

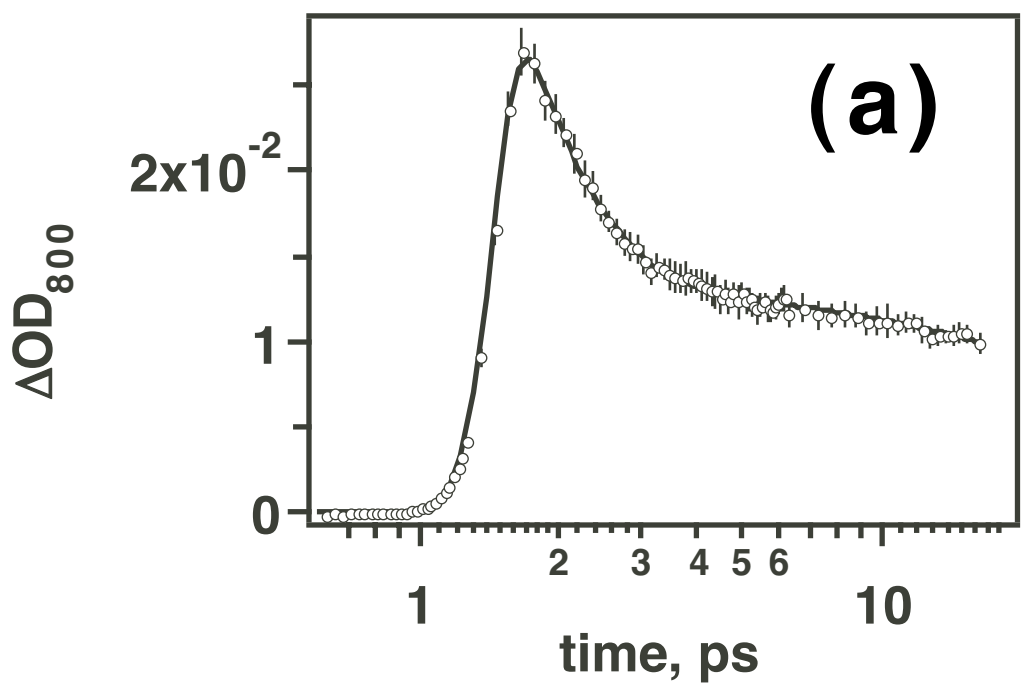
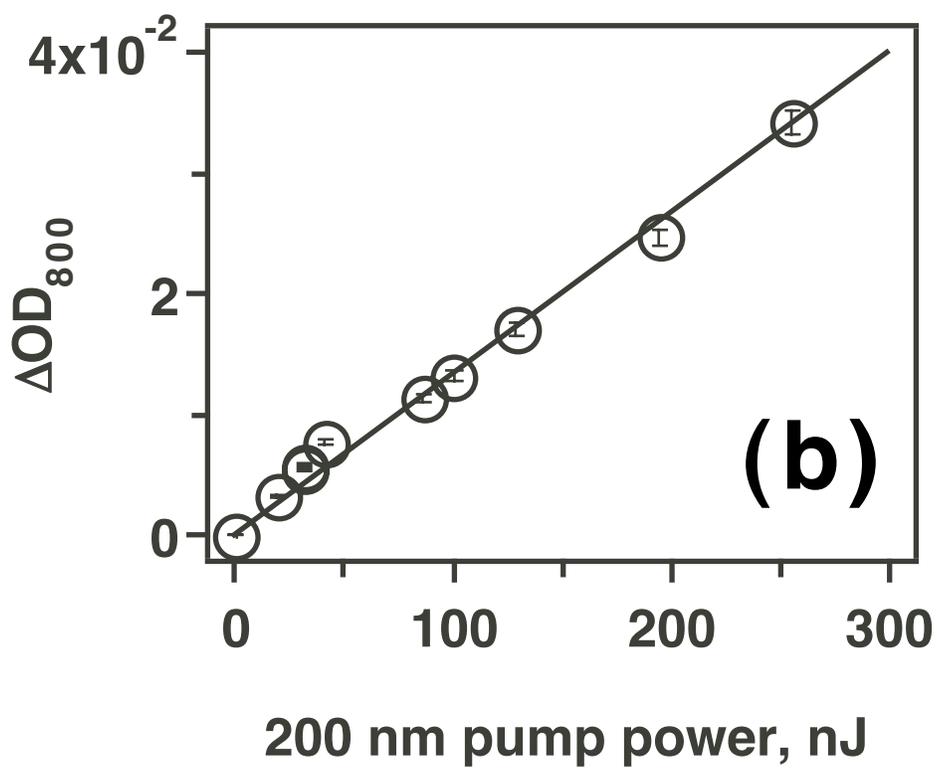

Lian et al.; Fig. 1S

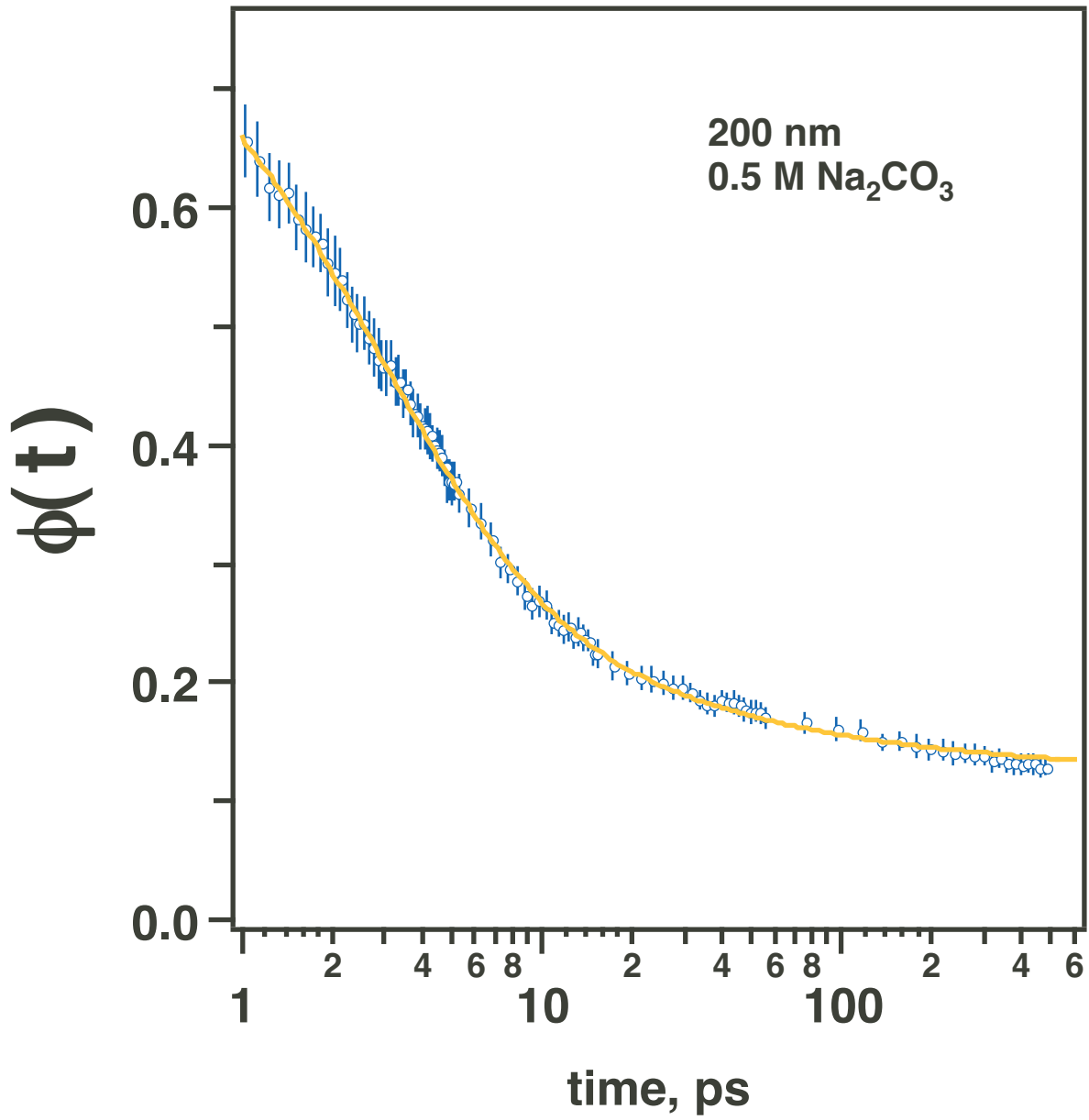

Lian et al.; Fig. 2S

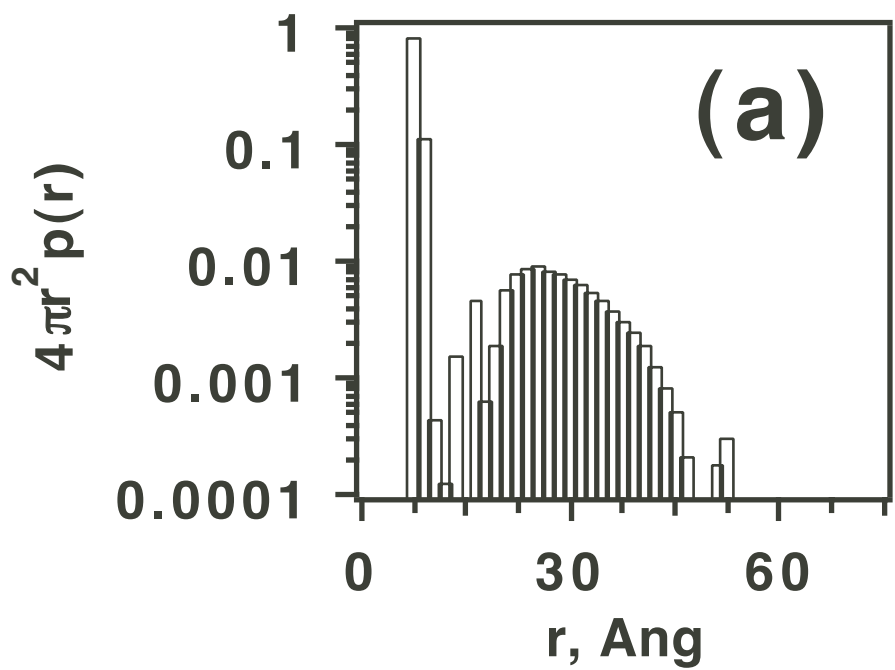

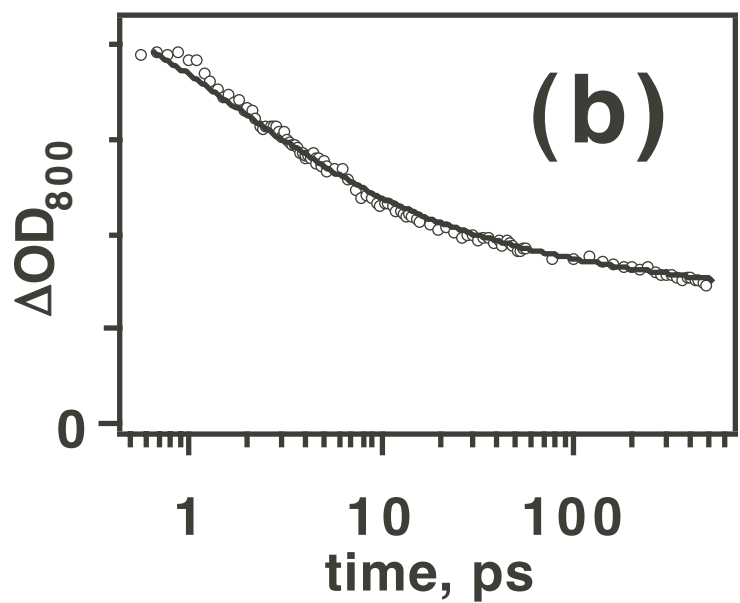

Lian et al.; Fig. 3S